\documentclass[%
reprint,
superscriptaddress,
amsmath,
amssymb,
aps,
prb,
floatfix,
]{revtex4-2}

\usepackage{graphicx}
\usepackage{adjustbox}
\usepackage{bm}
\usepackage[T1]{fontenc}
\usepackage[linktocpage=true,colorlinks=true,urlcolor=blue,linkcolor=red,citecolor = blue]{hyperref}
\usepackage[capitalize]{cleveref}
\usepackage{placeins}

\usepackage{xcolor}
\usepackage{soul}
\usepackage{comment}
\definecolor{applegreen}{rgb}{0.55, 0.71, 0.0}

\begin{document}

\title{Optimizing one-dimensional lithographically defined nanowires in semiconducting heterostructures}

\author{Leah Tom}
\thanks{These authors contributed equally to this work.}
\affiliation{Department of Physics, University of Wisconsin-Madison, Madison, WI 53706  USA}
\affiliation{Department of Electrical and Computer Engineering, University of California Los Angeles, Los Angeles, CA 90095  USA}

\author{Yik Kheng Lee}
\thanks{These authors contributed equally to this work.}
\affiliation{Chemical and Quantum Physics, School of Science, RMIT University, Melbourne, Australia}
\affiliation{ARC Centre of Excellence for Future Low-Energy Electronics Technologies, University of New South Wales, Sydney, New South Wales, 2052 Australia}

\author{Karina L.\ Hudson}
\affiliation{QuTech and Kavli Institute of Nanoscience, Delft University of Technology, Delft, The Netherlands}
\affiliation{ARC Centre of Excellence for Future Low-Energy Electronics Technologies, University of New South Wales, Sydney, New South Wales, 2052 Australia}

\author{Krittika Kumar}
\affiliation{School of Physics, University of New South Wales, Sydney, New South Wales 2052, Australia}
\affiliation{ARC Centre of Excellence for Future Low-Energy Electronics Technologies, University of New South Wales, Sydney, New South Wales, 2052 Australia}

\author{Jackson S.\ Smith}
\affiliation{Chemical and Quantum Physics, School of Science, RMIT University, Melbourne, Australia}
\affiliation{ARC Centre of Excellence for Future Low-Energy Electronics Technologies, University of New South Wales, Sydney, New South Wales, 2052 Australia}

\author{Alex R.\ Hamilton}
\affiliation{School of Physics, University of New South Wales, Sydney, New South Wales 2052, Australia}
\affiliation{ARC Centre of Excellence for Future Low-Energy Electronics Technologies, University of New South Wales, Sydney, New South Wales, 2052 Australia}

\author{S.\ N.\ Coppersmith}
\affiliation{School of Physics, University of New South Wales, Sydney, New South Wales 2052, Australia}
\affiliation{ARC Centre of Excellence for Future Low-Energy Electronics Technologies, University of New South Wales, Sydney, New South Wales, 2052 Australia}

\author{Jared H.\ Cole}
\affiliation{Chemical and Quantum Physics, School of Science, RMIT University, Melbourne, Australia}
\affiliation{ARC Centre of Excellence for Future Low-Energy Electronics Technologies, University of New South Wales, Sydney, New South Wales, 2052 Australia}

\author{Mark Friesen}
\affiliation{Department of Physics, University of Wisconsin-Madison, Madison, WI 53706  USA}
 
\begin{abstract}
One-dimensional nanowires can host remarkable electronic properties and are central to studying quantum transport. 
Although heterostructure-defined confinement combined with gate-defined potentials can greatly reduce the effects of impurities as measured via electron mobilities, the realization and interpretation of novel effects at this scale are complicated by disorder. 
Experiments on lithographically defined one-dimensional nanowires therefore do not exhibit ideal behavior. 
Using theoretical and computational models, we predict that optimizing the geometry of split gates in semiconducting heterostructures will lead to higher-quality channels for transport measurements. 
Further, we show that this optimization is essential for resolving features of interest such as the Rashba-induced spin gap. 
This new approach to lithographically defined conduction channels provides better energy resolution and is less susceptible to disorder.
\end{abstract}

\maketitle

\section{Introduction}
\label{Sec:Introduction}

    Quantum point contacts (QPCs) formed by surface gates on semiconductor heterostructures are used as quasi one-dimensional (1D) channels to study the role of quantum confinement in charge transport. 
    Examples include quantized conductance, the 0.7-anomaly and spin effects~\cite{Thornton1986, vanWees1988, Wharam1988, Thomas1996, Chen2009, Kumar2014, Micolich2011, Klochan2009}.
    They have also been used to probe the spin gap that emerges in the band structure of one-dimensional nanowires when Rashba and Zeeman interactions are both present~\cite{Quay2010, Heedt2017, Hudson2021}.
    Although modern GaAs/AlGaAs heterostructures with ultra-high mobility~\cite{Manfra2014} are ideal for observing such subtle effects, the resulting measurements of 1D conduction physics have often been underwhelming. 
    Limited reproducibility from device to device, relatively poor agreement with theoretical and computational models, and insufficient energy resolution in transport experiments have hindered our ability to identify features of interest.
    
    \begin{figure}[htp!] 
        \includegraphics[width=0.9\linewidth]{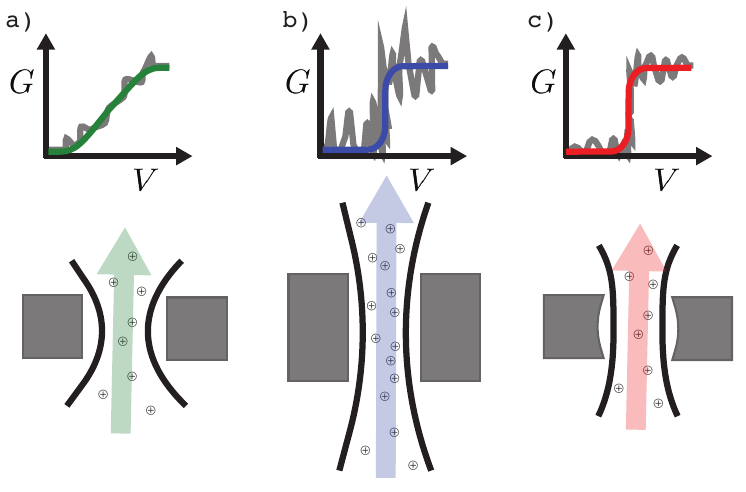}
        \centering
        \caption{
        Quantum point contacts formed within a 2DEG plane via metallic split gates display quantized conductance. 
        The (a) green, (b) blue, and (c) red lines show examples of ideal conductances ($G$) vs voltages ($V$) expected for (a) `short' and straight split gates, (b) `long' and straight split gates, and (c) elliptical split gates, respectively.
        In real devices, the presence of impurities results in noisy measurements as illustrated by the grey lines.
        Shorter elliptical gates can produce conductance steps that are as sharp as those produced by longer straight gates, but are less affected by impurities, demonstrating the benefit of optimizing the split-gate geometry.       
        }
        \label{fig:QPC_cartoon}
    \end{figure}

    A conventional QPC uses nanoscale metallic gates in a `split-gate' geometry to define a quasi-1D channel in the underlying confined charge distribution. 
    If the pinch-off region is narrow enough, conductance through the QPC is quantized and steps appear in conductance plots, as seen in Fig.~\ref{fig:QPC_cartoon}(a).    
    To resolve interesting spin effects using QPCs, the energy uncertainty in the channel needs to be smaller than the size of the effects.
    Therefore, the dwell time of the electrons in the channel should be maximized, as implied by the energy-time uncertainty principle ($\Delta E \Delta t \geq \hbar/2$).     
    An intuitive way to do this would be to increase the length of the channel.
    This should make the conductance steps sharper~\cite{Buettiker1990} and allow finer features to be resolved. 
    However, a longer conduction channel also allows the charge carriers to interact with more impurities, which results in noisy measurements that can obscure these features, as seen in Fig.~\ref{fig:QPC_cartoon}(b). 

    \begin{figure}[htbp]
        \centering
        \includegraphics[width=0.9\linewidth]{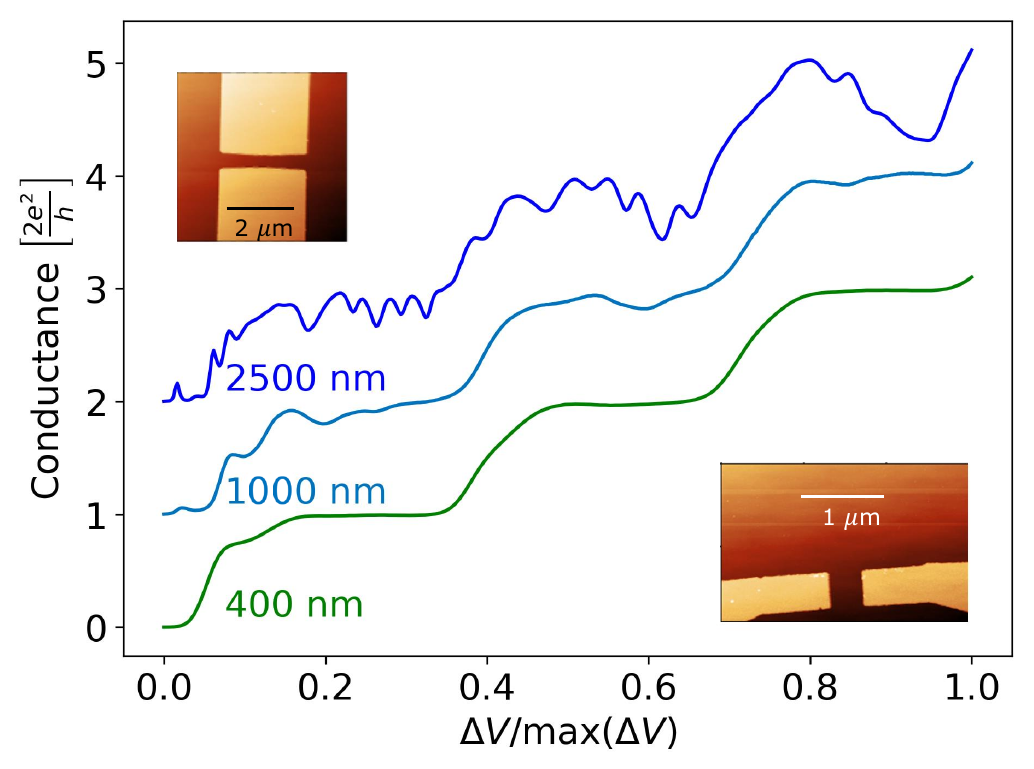}
        \caption[Conductance data of electron QPCs at different lithographic lengths with Straight Gate Design] {
        Experimental conductance data measured for QPCs formed by rectangular split gates of lengths 400~nm, 1000~nm, and 2500~nm. 
        The gate separation was fixed at 400~nm for all gate lengths.
        The source-drain bias was set to 100~{\textmu}V and the measurement temperature was 250~mK.
        $\Delta V$ is the difference between the applied gate voltage and the pinch-off voltage (the voltage when current stops flowing through the QPC).
        For each trace, the voltage range is adjusted to include approximately 3.1 conductance steps, $\Delta V$ is normalized by the maximum value of the given trace, $\mathrm{max}(\Delta V)$, and the traces are offset for clarity.
        The inset AFM (atomic force microscope) images show examples of split gates of lengths 400~nm (bottom right) and 2500~nm (top left), similar to those used for these measurements.
        It can be seen that as gate length increases, the conductance plateaus become less well-defined as the paths of the electrons are affected by more impurities.
        }
        \label{fig:qpc_electron_data}
    \end{figure}
    
    Figure~\ref{fig:qpc_electron_data} shows an experimental example of conductance data for rectangular split-gate QPCs with lithographic lengths of 400~nm, 1~{\textmu}m, and 2.5~{\textmu}m.
    The devices were fabricated and measured with similar methods to those used by Hudson et al.~\cite{Hudson2021}, which are summarized in Appendix~\ref{appendix_methods}.
    Quantized conductance plateaus are observed, providing evidence of quasi-ballistic transport, which is necessary to precisely control the number of conducting modes. 
    However, the devices with longer gates exhibit less well-defined conductance plateaus than the devices with shorter gates due to interaction with more impurities within the channel region of the QPC.

    In this paper, we use a combination of theoretical and computational models to show that the quality of 1D channels can be substantially improved by changing the shape of the surface gates to extend the region over which the effective width of the channel is approximately uniform, as illustrated in Fig.~\ref{fig:QPC_cartoon}(c).
    This results in channels without bottlenecks that are particularly susceptible to fluctuations caused by trapped charge.
    The paper is organized as follows.
    Section~\ref{Sec:Comsol potentials} introduces the device geometry and presents the simulations used to obtain the electrostatic potential in the two-dimensional electron gas (2DEG) with an optimized gate geometry.
    Section~\ref{Sec:Conductance_calculations} then presents transport calculations, which show that replacing straight split gates with elliptical split gates can significantly improve the quality of the resulting 1D channel.
    In Sec.~\ref{Sec:saddle}, we briefly summarize a simple analytical model of conductance through a saddle-shaped confinement potential.
    In Sec.~\ref{Sec:gate_type}, we examine conductance spectra for QPCs formed in gates of varying length, using the results of the detailed electrostatics simulations.
    In Sec.~\ref{Sec:disorder}, we study the impacts of disorder in the same devices.
    In Sec.~\ref{Sec:spin_orbit}, we show how the new gate geometries could improve experiments aimed at resolving the Rashba spin gap.
    We finally conclude in Sec.~\ref{Sec:Summary}.

\section{Design considerations for Quantum Point Contacts}
    \label{Sec:Comsol potentials}
    
    \begin{figure}[htbp]
        \centering
        \includegraphics[width=0.9\linewidth]{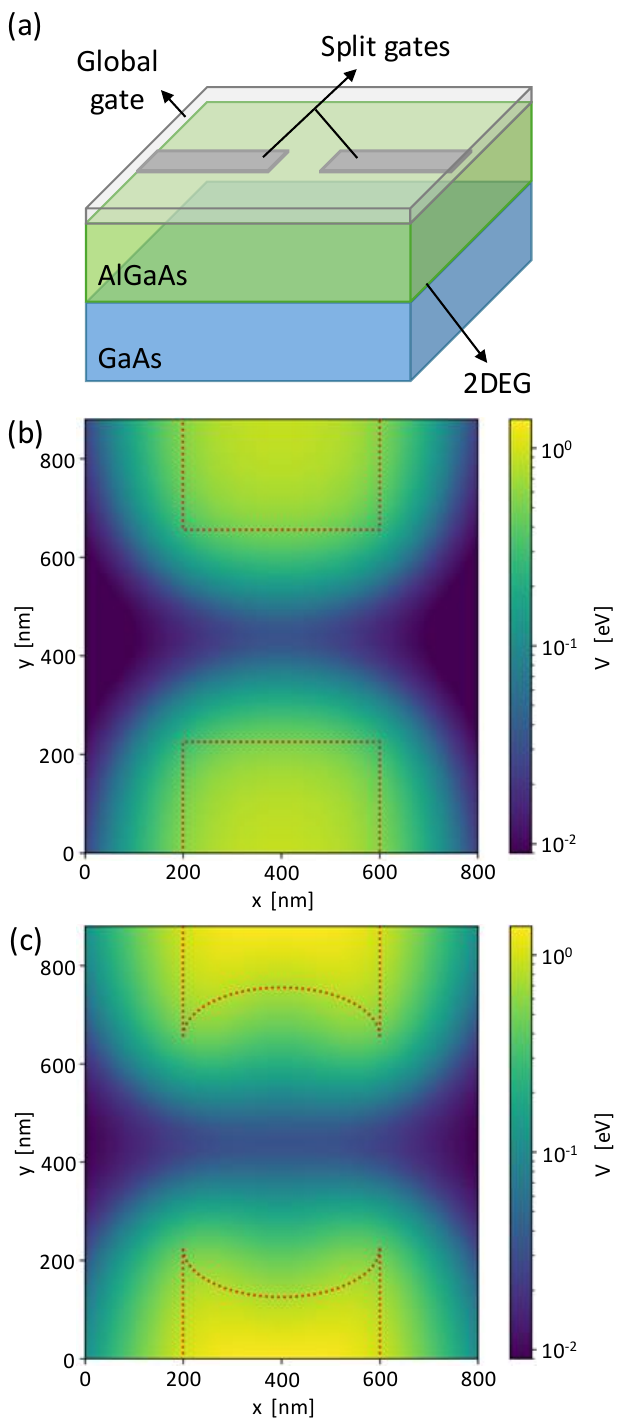}\\
        \caption[Typical electrostatics simulations.]
        {Electrostatics simulations.
        (a) Schematic diagram of the GaAs/AlGaAs heterostructure.
        A thin, conformal oxide layer separates the split gates from a global top gate. 
        A positive bias on the global gate accumulates electrons in the 2DEG, formed at the GaAs/AlGaAs interface, while a negative bias on the split gates locally depletes the electrons below, resulting in a quasi-1D channel.
        (b,c) Color maps of the simulated electrostatic potentials $V(x,y)$ (log scale), obtained in the 2DEG plane, for devices with 400~nm lithographic lengths.
        Two different gate geometries are considered in this work, defined by (b) `straight' split gates, or (c) `elliptical' split gates.
        Red dotted lines indicate the outlines of the split gates above the 2DEG.
        }
        \label{fig:experimental_setup_qpc}
    \end{figure}
    
    The device setup in our simulations is shown schematically in Fig.~\ref{fig:experimental_setup_qpc}(a), to match the experiments described in Fig.~\ref{fig:qpc_electron_data}.
    As described in Appendix~\ref{appendix_methods}, metal split gates and a global accumulation gate are deposited on top of an GaAs/AlGaAs heterostructure.
    The gates are electrically isolated from one another by an oxide layer. 
    The 2DEG in this system forms at the GaAs/AlGaAs interface, due to the intrinsic conduction-band offset. 
    A uniform positive bias is applied to the accumulation gate to control the electron density in the 2DEG. 
    A negative bias is applied to the split gates, to modulate the 2DEG below, such that it forms a quasi-1D channel.
    This structure is modeled using COMSOL software \cite{COMSOL}, as described in Appendix~\ref{appendix_Leah}, to determine the full three-dimensional (3D) electrostatic potential, from which we extract the confinement potential in the plane of the 2DEG.
    This potential is then used in the transport calculations that follow.

    We consider two types of split-gate geometries.
    The first is a `straight' gate, which has rectangular inner edges, as outlined in Fig.~\ref{fig:experimental_setup_qpc}(b).
    The second is an `elliptical' gate, which has elliptical inner edges, as illustrated in Fig.~\ref{fig:experimental_setup_qpc}(c). 
    The goal of the elliptical geometry is to reduce the curvature of the electrostatic potential around its saddle point, thereby enabling a straighter, more-uniform transport channel. 
    We define a `quasi-1D' channel as a region of the device where just a single subband mode is occupied, due to its transverse confinement energy lying below the Fermi level.
    In Appendix~\ref{appendix_Leah}, we show that significantly longer quasi-1D channels can be obtained by employing elliptical gate patterns.
    
    The lithographic lengths of the split gates are defined along the transport direction.
    For example, the gates in Figs.~\ref{fig:experimental_setup_qpc}(b) and \ref{fig:experimental_setup_qpc}(c) are both 400~nm long.
    The separation between the closest points on the split gates is fixed at 430~nm for all simulations.
    Figures~\ref{fig:experimental_setup_qpc}(b) and \ref{fig:experimental_setup_qpc}(c) also show color maps of the electrostatic potentials evaluated in the plane of the 2DEG, for the gate geometries shown.
    For the rectangular gate in Fig.~\ref{fig:experimental_setup_qpc}(b), we observe an electrostatic channel whose width is constant only near the center, flaring out towards the reservoir regions of the 2DEG on either side.
    In contrast, the effective channel for the elliptical gate in Fig.~\ref{fig:experimental_setup_qpc}(c) exhibits a more-uniform width, over a greater distance, despite having the same lithographic length.
    In the following sections we study how this improved channel uniformity affects its conductance.

\section{Conduction in quantum point contacts}
    \label{Sec:Conductance_calculations}
    
    We now perform ballistic transport simulations of QPC devices defined by the straight and elliptical gate geometries described in Sec.~\ref{Sec:Comsol potentials}.
    Transport calculations were performed within the non-equilibrium Green's function formalism using Green's functions and lead self-energies obtained from the Kwant software package (version 1.5.0) \cite{Datta2012,Groth2014,Lee2022}.
    For each device geometry, we take a horizontal slice through the 3D electrostatic potential, obtained from the COMSOL simulations, at a depth of 110~nm below the surface gates where the 2DEG resides, yielding a 2D confinement potential $V(x,y)$, which is then used in the Kwant simulations.

    The full 2D quantum mechanical model includes Zeeman and Rashba spin-orbit coupling, as described by the following Hamiltonian~\cite{Usaj2004}:
    \begin{multline}
        H = \frac{\hbar^2}{2m^{*}}\left(k^{2}_{x}+k^{2}_{y}\right)+V(x,y) \\
        +\frac{g}{2}\mu_{B}\sigma^{}_{z}B^{}_{z}+\alpha\left(k_{y}\sigma_{x}-k_{x}\sigma_{y}\right) ,
        \label{Eq:Hamiltonian}
    \end{multline}
    where $m^{*}=0.067m_e$ is the effective mass of an electron in GaAs~\cite{Winkler2003}, $\mathbf{k}=(k_x,k_y)$ is the in-plane wavevector, $g=-0.44$ is the effective $g$-factor, $\mu^{}_{B}$ is the Bohr magneton, $\{\sigma_{i}\}$ are Pauli spin matrices, $B_{z}$ is the magnitude of the out-of-plane magnetic field, and $\alpha$ is the Rashba coupling parameter (defined below).
    The effect of an external magnetic field on the orbital physics was included via the Peierls substitution~\cite{Peierls1933a}:
    \begin{equation}
        \mathbf{k} \Rightarrow \mathbf{k}-\frac{e}{\hbar}\mathbf{A}
        \label{Eq:Peierls}
    \end{equation}
    where -$e$ is the charge of an electron, $\hbar$ is the reduced Planck's constant, and $\mathbf{A}$ is the magnetic vector potential.

    Conductance was calculated using the Landauer-B\"uttiker formalism, with a vanishing bias voltage, given by \cite{Datta1995}:
    \begin{align}
        \label{Eq:conductance}
        G &= \frac{2e^{2}}{h}\int T(E)\left[-\frac{\partial f_{0}(E)}{\partial E}\right]dE ,
    \end{align}
    where $T(E)$ is the transmission probability and $f_{0}(E)$ is the Fermi distribution function.
    Here, we take the zero-temperature limit, for which the derivative in Eq.~(\ref{Eq:conductance}) reduces to a Dirac delta function, such that the conductance is proportional to the transmission probability.
    We then calculate the conductance through the QPCs as a function of energy, focusing on the energy range starting at zero and including two conductance plateaus. 

\subsection{Simplified conductance model}
\label{Sec:saddle}
       
In the Kwant simulations described below, we include the full electrostatic potential maps obtained in Sec.~\ref{Sec:Comsol potentials}.
To help analyze these simulation results, we first consider a simplified model of a confinement potential, which allows approximate analytical solutions.
This is the classic problem of a saddle-shaped confinement potential \cite{Buettiker1990}.
The results obtained from this model provide a convenient framework for analyzing the numerical solutions.
Indeed, we find that our more-detailed solutions described below -- which include the full electrostatic potential but exclude disorder -- are well-described by this simple theory.
We do not re-derive the saddle-potential theory here, but simply summarize the main results.

The shape of the confinement potential formed by the split gates can be locally approximated by a saddle-shaped confinement potential, given by
    \begin{equation}
        V(x,y) = V_0 - \frac12 m^* \omega_x^2 x^2 + \frac12 m^* \omega_y^2 y^2 ,
        \label{eq:Vxy}
    \end{equation}
    where $V_0$ is the electrostatic potential at the saddle point, and $\omega_x$ and $\omega_y$ parameterize the shape of the saddle. 
    The Hamiltonian from Eq.~(\ref{Eq:Hamiltonian}) can be solved using Eq.~(\ref{eq:Vxy}), in the limit of $g=\alpha=0$, obtaining the energy eigenvalues
    \begin{equation}
        \epsilon_n = \frac{2}{\hbar \omega_x}\left[E-\hbar \omega_y \left(n + \frac12 \right) - V_0 \right] ,
    \end{equation}
    where $n=0,1,\dots$ labels the mode index.
    Note that we have offset the total energy $E$ by an appropriate subband energy, and we have made the energy dimensionless, in terms of the longitudinal energy scale $\hbar\omega_x/2$.
    The ballistic transmission probabilities, from the $n^\text{th}$ incident channel to the $m^\text{th}$ outgoing channel, can then be computed as
    \begin{equation}\label{Eq:transmissionfunction}
        T_{mn} = \delta_{mn} \frac{1}{1+e^{-\pi \epsilon_n}} ,
    \end{equation}
    where the transport is assumed to be ballistic.
    
    In the simulations that follow, for systems without spin-orbit coupling or disorder, we fit our numerical results to Eq.~(\ref{Eq:transmissionfunction}), allowing us to obtain an estimate for the ratio $\omega_y/\omega_x$. 
    According to the standard theory \cite{Buettiker1990}, this ratio allows us to quantify the `sharpness' of the steps between conductance plateaus, with sharper steps corresponding to larger values of $\omega_y/\omega_x$.
    As noted above, sharp steps are important for achieving better energy resolution when performing QPC spectroscopy, and more specifically, for resolving subtle spin effects.

    \subsection{Effect of gate geometry on conductance}
        \label{Sec:gate_type}

        \begin{figure}[htbp]
            \centering
            \includegraphics[width=0.9\linewidth]{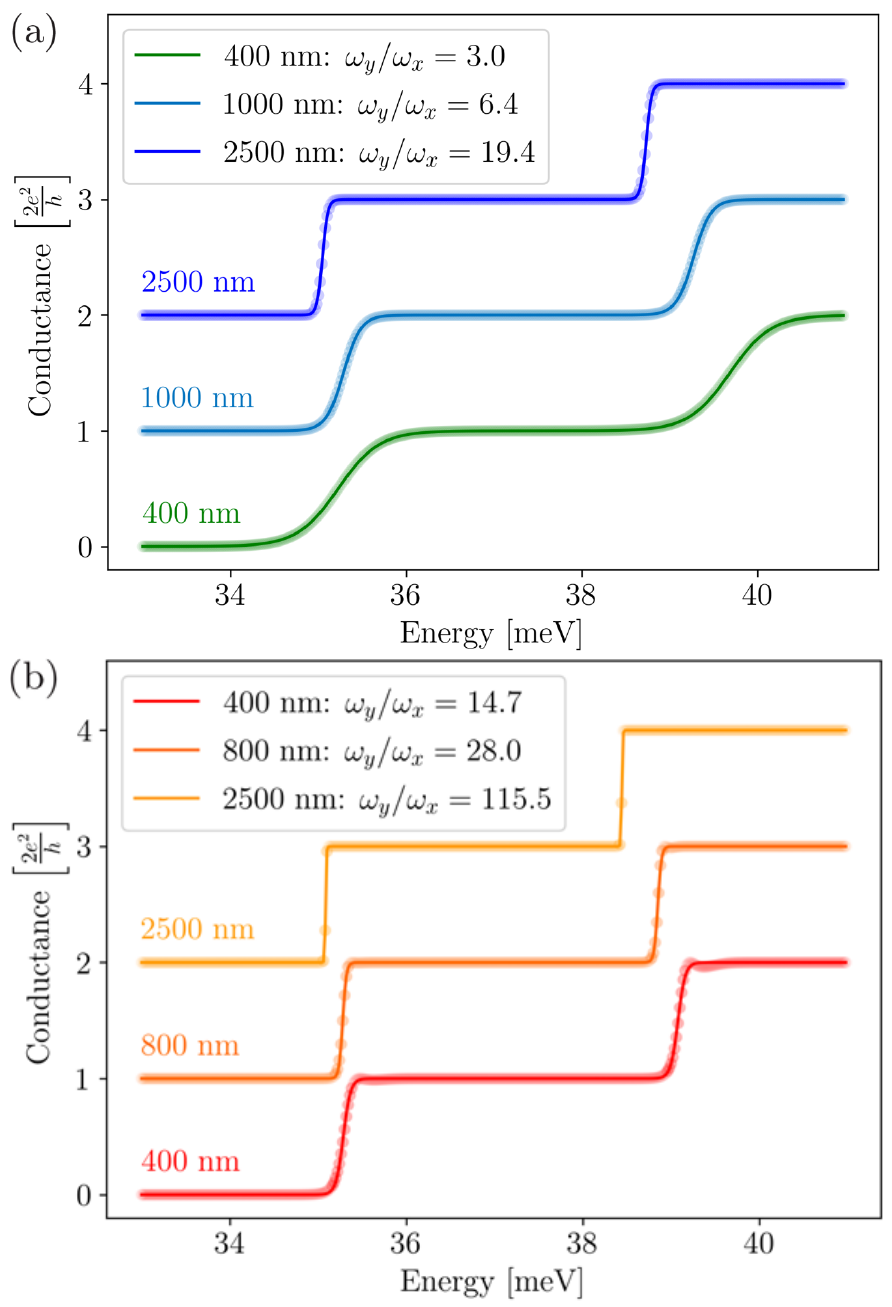}\\
            \caption[Conductance for different gate shapes] {
            Conductance versus energy simulation results.
            (a) Results for straight split gates with lithographic lengths of 400~nm, 1000~nm, and 2500~nm.
            (b) Results for elliptical split gates with lithographic lengths of 400~nm, 800~nm, and 2500~nm.
            Here, the minimum gate separation is given by 430~nm for all devices, and the magnetic field strength and Rashba coupling are set to zero.
            Conductance traces for the different geometries are offset for clarity. 
            Numerical results of the conductance calculations are shown as light-colored circles.
            The simulation data are also fit to Eq.~(\ref{Eq:transmissionfunction}), as shown by the dark-colored solid lines.
            The fitting parameters $\omega_y/\omega_x$ are indicated in the legends, demonstrating that elliptical gates can yield substantially sharper conductance steps than straight gates.
            }
            \label{fig:Conductance_var_lengths}
        \end{figure}
        
        The results of transport simulations for the conductance through a QPC formed by straight split gates, with lithographic lengths of 400~nm, 1000~nm, and 2500~nm, are shown in Fig.~\ref{fig:Conductance_var_lengths}(a).
        For simplicity, we again exclude Zeeman and Rashba effects here, by setting $g=\alpha=0$ in Eq.~(\ref{Eq:Hamiltonian}).
        As noted above, the slopes of the steps between the conductance plateaus provide important information regarding energy resolution. 
        In the figure, we see that the steps between the conductance plateaus are somewhat sloped for the 400~nm long gates, gradually becoming steeper as the length of the split gates increases.
        At 2500~nm, the steps are almost vertical, approaching the perfect step function that one would expect from the conductance plot of an idealized nanowire with uniform width. 
        Each of these curves is fit to Eq.~(\ref{Eq:transmissionfunction}), with results shown as solid lines in Fig.~\ref{fig:Conductance_var_lengths}(a), and the resulting fitting parameters shown in the legend.
        We note that the fitting function works very well for this data, indicating a systematic increase in the ratio $\omega_y/\omega_x$ as a function of the length of the split gates.

        The results of transport simulations for conductance through a QPC formed by elliptical split gates, with the same lithographic lengths of 400~nm, 800~nm, and 2500~nm, are shown in Fig.~\ref{fig:Conductance_var_lengths}(b).
        A comparison with Fig.~\ref{fig:Conductance_var_lengths}(a) shows that, for equal gate lengths, the elliptical split gates produce much steeper step edges (i.e., larger $\omega_y/\omega_x$) between the conductance plateaus than straight split gates.
        Therefore, longer elliptical split gates should provide better energy resolution for resolving features in the conductance spectrum.
        In the next subsection, we consider disorder and show that simply increasing the length of the gates is not always the optimal solution.

    \subsection{Effect of disorder on conductance}
        \label{Sec:disorder}
         
        \begin{figure}[t]
            \includegraphics[width=0.9\linewidth]{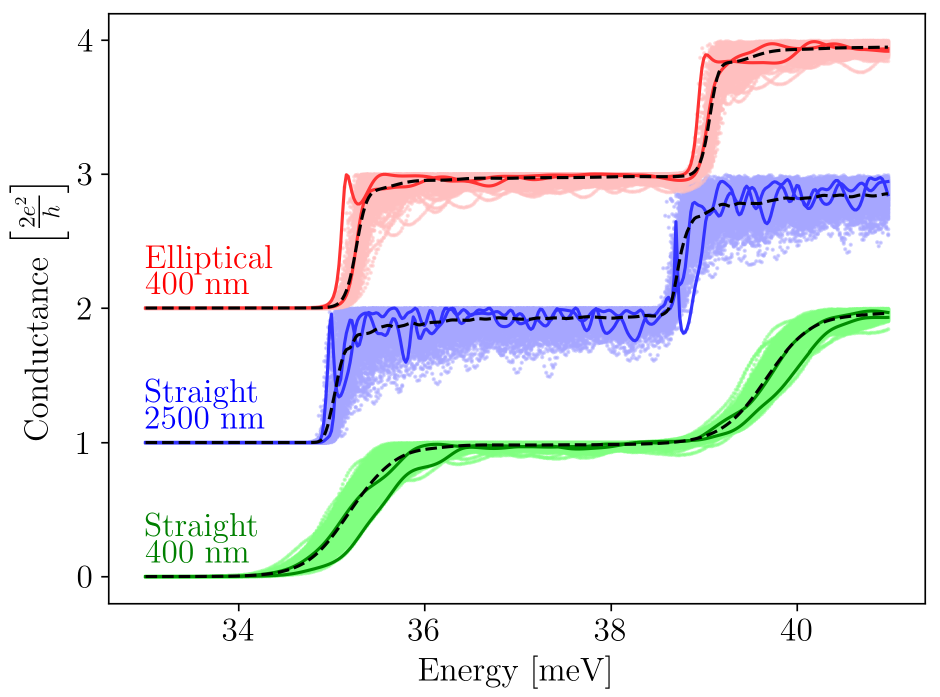}
            \caption{
            Conductance versus energy simulation results, calculated for QPCs formed by straight split gates of lengths 400~nm (green) and 2500~nm (blue), and elliptical gates of length 400~nm (red).
            Here, the minimum gate separation is given by 430~nm for all devices, and the magnetic field strength and Rashba coupling are set to zero.
            Disorder is included in the simulations using a simple Anderson model.
            For each gate geometry, we consider 200 independent realizations of random on-site disorder, with results plotted as light-colored circles.
            Data for each geometry is vertically offset for clarity. 
            The black dashed lines show mean conductances of all 200 disorder realizations, for each gate type, while the darker-colored solid lines show conductances for realizations with the highest and lowest root-mean-squared deviations from the mean conductance curves.
            The figure shows that, in the presence of disorder, elliptical split gates of length 400~nm yield more well-defined conductance steps than straight split gates of length 2500~nm.
                }
                \label{fig:Conductance_disorder}
        \end{figure}
        
        We now briefly explore the effects of disorder on the conductance of QPCs formed by straight and elliptical split gates.        
        Disorder is modeled in our simulations by randomly varying the electrostatic potential in the 2DEG~\cite{Lee2022}, to account for the effects of trapped charges in the gate stack \cite{Connors2019_chargenoise}, imperfections in the crystal lattice \cite{Choi2013_nativepointdefectAl2O3, Holder2013_HydrogenDefectsAlumina, Weber2011_NativeDefectsMOSdevices,Goes2008}, or other defects. 
        We apply a simple Anderson model~\cite{Anderson1958}, for which independent random potentials are drawn from a uniform distribution in the range \(\left[ - \frac{U_{0}}{2},\frac{U_{0}}{2} \right]\) and added to every grid point.
        For the calculations in this subsection, we adopt a value of approximately 3.6~meV for $U_{0}$, which corresponds to a mean free path of about 7~{\textmu}m~\cite{Lee2022} when the Fermi energy is equal to 35~meV.
        This mean free path is larger than the device dimensions, so the transport is still ballistic.
        As in the previous subsection, we again set $g=\alpha=0$.
        
        While keeping $U_{0}$ constant, we generate 200 independent realizations of the random on-site disorder and calculate the conductance for QPCs formed by straight split gates of length 400~nm and 2500~nm, and for elliptical split gates of length 400~nm.
        The results of the individual disorder realizations are shown as lighter-colored dots in Fig.~\ref{fig:Conductance_disorder}, and the corresponding mean conductances are plotted as black dashed lines.
        The conductances for the individual disorder realizations with the highest and lowest root-mean-squared deviations from the corresponding means are also plotted as dark-colored solid lines, to highlight the device-to-device variations that would be seen in experimental measurements.
        Overall, the behavior is fairly consistent with the experimental results shown in Fig.~\ref{fig:qpc_electron_data}.

Taking a closer look, we observe some expected behavior: for example, the mean conductances are seen to mirror their nondisordered counterparts in Fig.~\ref{fig:Conductance_var_lengths}, albeit with some suppression of the conductance near the step edges, making their energies more difficult to resolve.
By comparing the longer gates to the shorter gates, we observe some resilience of the latter to charge fluctuations; this can be seen both in the narrower spread of the conductance fluctuations for shorter gates and the overall suppression of extreme fluctuation events.
Such behaviors are reasonable given that, for a fixed mean free path, electrons have a higher chance of scattering when traveling through a longer channel.
Moreover, increasing the length of the channel increases the chances of encountering a defect that could render the device nonfunctional. 
Finally, we compare the conductance in shorter elliptical gates to longer straight gates, observing similarly sharp steps, with significantly suppressed disorder for the elliptical gates.
This has important ramifications for reproducibility and device yield in disordered devices, strongly advocating for the use of elliptical gates.

    \subsection{Resolving Rashba-induced spin splitting}
        \label{Sec:spin_orbit}
        
        \begin{figure}[htbp]
            \centering
            \includegraphics[width=0.9\linewidth]{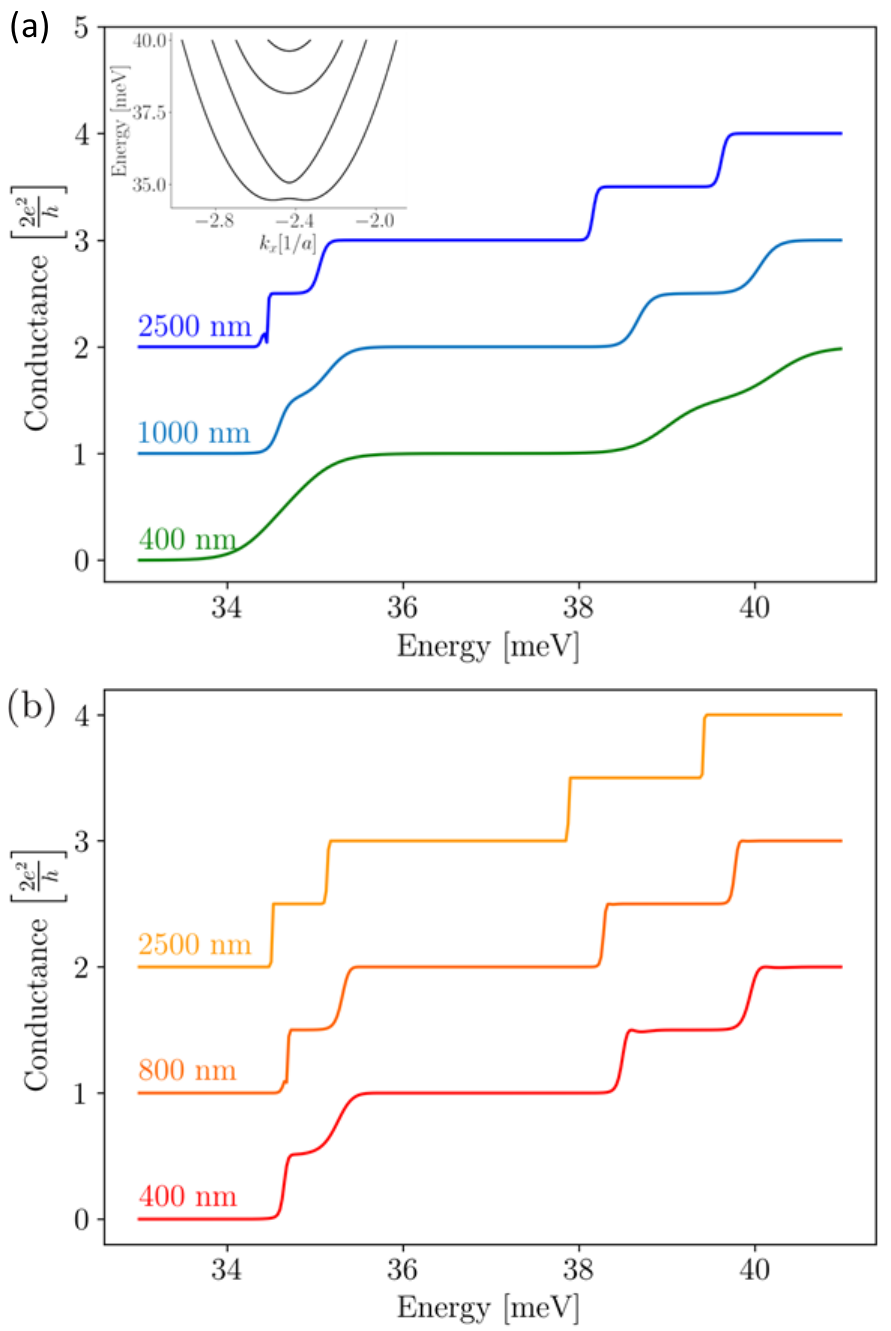}\\
            \caption{
            Conductance versus energy simulation results, including the effects of Rashba spin-orbit coupling, with a  coupling parameter of $3.1 \times 10^{-11}$~eV$\cdot$m, and an applied magnetic field of 1~T.
            (a) Results for straight split gates with lithographic lengths of 400~nm, 1000~nm, and 2500~nm.
            (b) Results for elliptical split gates with lithographic lengths of 400~nm, 800~nm, and 2500~nm.
            All other device parameters are the same as Fig.~\ref{fig:Conductance_var_lengths}.
            The results are offset vertically for clarity.
            The inset in (a) shows the band structure calculated for an `ideal' 1D wire, as explained in the main text.
            Here, the spin-gap feature can only be observed at the very bottom of the lowest subband, requiring a level of energy resolution that is not consistent with our simulations.
            Although the characteristic dip associated with a spin gap is not observed in the conductance spectra here, the widths of the conductance steps accurately reflect the energy gaps in the band structure shown in the inset.
            These results demonstrate that the half-quantum spin-splitting is well-defined for the case of elliptical split gates, even for shorter gates, in contrast with straight-gate geometries.
            }
            \label{fig:Conductance_Rashba}
        \end{figure}
        
        When an out-of-plane magnetic field is applied to the 2DEG, the Zeeman effect lifts the spin degeneracy of the energy bands, causing conductance plateaus to appear at half-integer multiples of the conductance quantum, rather than integer multiples.
        When combined with the Rashba effect, a `spin gap' is opened in the energy spectrum, which can potentially be observed as a dip of half a conductance quantum \cite{Quay2010, Heedt2017, Kammhuber2017, Hudson2021}, when the conductance spectrum has sufficient energy resolution.
        However, since the spin gap appears at energies very near the step edges in the conductance spectra, the dip feature may be obscured if the step edge is not well-defined.
        Due to the improved energy resolution afforded by elliptical gate structures, we now explore whether these devices could provide sufficient detail in the conductance data to observe well-defined spin-orbit features.

        To answer this question in our transport simulations, we explore the emergence of spin-split conductance steps when the magnetic field and Rashba spin-orbit coupling terms are included in Eq.~(\ref{Eq:Hamiltonian}).
        We therefore repeat the conductance simulations of Fig.~\ref{fig:Conductance_var_lengths}, for the same gate geometries, while now including an out-of-plane magnetic field of 1~T.
        Since the Rashba spin-orbit coupling for electrons in GaAs is rather weak, we consider here an elevated coupling parameter of $3.1 \times 10^{-11}$~eV$\cdot$m, which could potentially be engineered by applying a large vertical electric field \cite{Eldridge2010} or by appropriate doping \cite{Huang2011, Lo2017}.
        
        The results of these transport simulations are presented in Fig.~\ref{fig:Conductance_Rashba}.
        Figure~\ref{fig:Conductance_Rashba}(a) shows results for straight-gate structures; here, we note that the first spin-split conductance step is fully obscured for the 400~nm device, although it resolves as the gate length increases.
        In contrast, for elliptical gates, spin-split steps are observed for all the gate lengths shown in Fig.~\ref{fig:Conductance_Rashba}(b); moreover, the resolution of the energy splitting between the steps becomes sharper for longer gates.

We now comment briefly on the spin-orbit features observed in our simulations, particularly in relation to the anticipated dip feature in the conductance spectrum, caused by the spin gap.
First, to illustrate the emergence of a spin gap in the QPC devices studied here, we compute the dispersion relation of an `ideal' wire.
For our purposes, we define an ideal wire as an infinite, uniform 1D wire whose confinement potential is given by a linecut in the 2D electrostatic potential, passing transversely through the center of the 2500~nm straight QPC.
Using this confinement potential, we solve Eq.~(\ref{Eq:Hamiltonian}) to obtain the dispersion relation shown in the inset of Fig.~\ref{fig:Conductance_Rashba}(a).
The spin gap is potentially observed as a dip in the conductance spectrum, which occurs when the number of conductance channels is suddenly reduced \cite{Quay2010}.
For the dispersion relation in Fig.~\ref{fig:Conductance_Rashba}(a), we see that an energy resolution of about 40~{\textmu}eV would be needed to observe such a gap feature, even in the ideal case; however, we see that none of the conductance traces plotted in Fig.~\ref{fig:Conductance_Rashba} has this level of resolution, and no dips are observed in the spectra. 
Moreover, when disorder effects are included in the simulations, the resolution is further reduced.
Thus, we would not expect to observe the anticipated conductance dips in these devices.
We note that this is a reflection of the materials system that we consider, not the gate geometries, which are the focus of this work.

On the other hand, we note that the half-quantum, spin-split conductance steps are readily observed, and should also survive the inclusion of disorder in the simulations.
We further note that the emergence of these spin-split steps is a clear signature of spin-orbit effects.
Indeed, the spin splitting caused by the Zeeman term alone, i.e. without including the Rashba term in the Hamiltonian, is only $\sim$25~{\textmu}eV for the 1~T magnetic field we consider.
The observed energy splitting of the spin-split conductance steps in Fig.~\ref{fig:Conductance_Rashba} is clearly larger than this Zeeman splitting, and must therefore include spin-orbit effects.

To summarize this subsection, we find that Rashba-induced energy splittings can only be resolved using very long (micron-scale) gate structures, for the case of straight-gate QPCs.
However, similar results can be achieved with much shorter elliptical split gates. 
The improved performance is especially important when considering charge disorder, which can obscure the spectroscopic features, as illustrated in Fig.~\ref{fig:Conductance_disorder}.

\section{Conclusion}
    \label{Sec:Summary}
    
    In summary, we have used a combination of simulation tools to study the role of the lithographic gate geometry on the performance of QPCs formed between pairs of split gates.
    We find that elliptical gates are preferable to rectangular gates for their ability to provide sharp step edges in a series of conductance plateaus, as a function of gate voltage. 
    Since step sharpness determines the energy resolution of conductance spectroscopy measurements, this suggests that elliptical gates should outperform straight gates in such applications.
    This is especially important when disorder is present, as it is in all devices, since shorter elliptical gates are more resilient to charge noise than longer rectangular gates.
    Indeed, we show that straight-gate simulations, including disorder, are largely consistent with experimental measurements.
    We have further demonstrated that sharp step edges are essential for observing subtle features in the conductance spectrum, arising from spin-orbit effects, such as half-quantum conductance steps and spin gaps. 
    These results point to a new approach to lithographically defined conduction channels that provides better energy resolution and is less susceptible to disorder.

\section*{Acknowledgments}
    This work was supported in part by the Australian Research Council Centre of Excellence in Future Low-Energy Electronics Technologies (FLEET), Project No.\ CE170100039, funded by the Australian government.
    It was also funded through Award No.\ \#DE-SC0020313, by the U.S. Department of Energy, Office of Science, and by the Australian Research Council Discovery Project DP210101608.
    Finally, it was supported in part by an unrestricted gift to UNSW Sydney from Google AI.
    Computational resources and services were provided by the National Computational Infrastructure (NCI), which is supported by the Australian Government.

\section*{Data Availability}
The COMSOL \cite{COMSOL} and Kwant \cite{Groth2014} simulation code used in this work, along with the raw data in the figures, are available on Zenodo \cite{zenodo}.

\appendix

\section{Experimental methods used in Fig.~\ref{fig:qpc_electron_data}}
\label{appendix_methods}

The (001) GaAs/Al$_x$Ga$_{1-x}$As heterostructures used in this work were grown by molecular beam epitaxy (MBE). 
The 2DEG was formed at the hetero-interface located 110~nm below the surface. 
The structures were undoped and operated in accumulation mode.
The split gate devices were fabricated using standard electron beam lithography techniques on a metal film formed by thermally evaporating a 12~nm Ti sticking layer followed by a 30~nm layer of Au. 
Ohmic contacts to the 2DEG were made by thermally evaporating a Ni/AuGe/Ni stack into etched pits, followed by an annealing step. 
Finally, a 20~nm Al$_2$O$_3$ global dielectric layer was applied by atomic layer deposition and a global electrode was defined using photolithography and thermal evaporation of 20~nm of Ti and 80~nm of Au. 
QPC transport measurements were performed using standard lock-in techniques with an ac excitation voltage of 100~{\textmu}V at 250~mK.

\section{COMSOL simulations of the QPCs}
    \label{appendix_Leah}

    \begin{figure}[t!]
            \centering
            \includegraphics[width=0.75\linewidth]{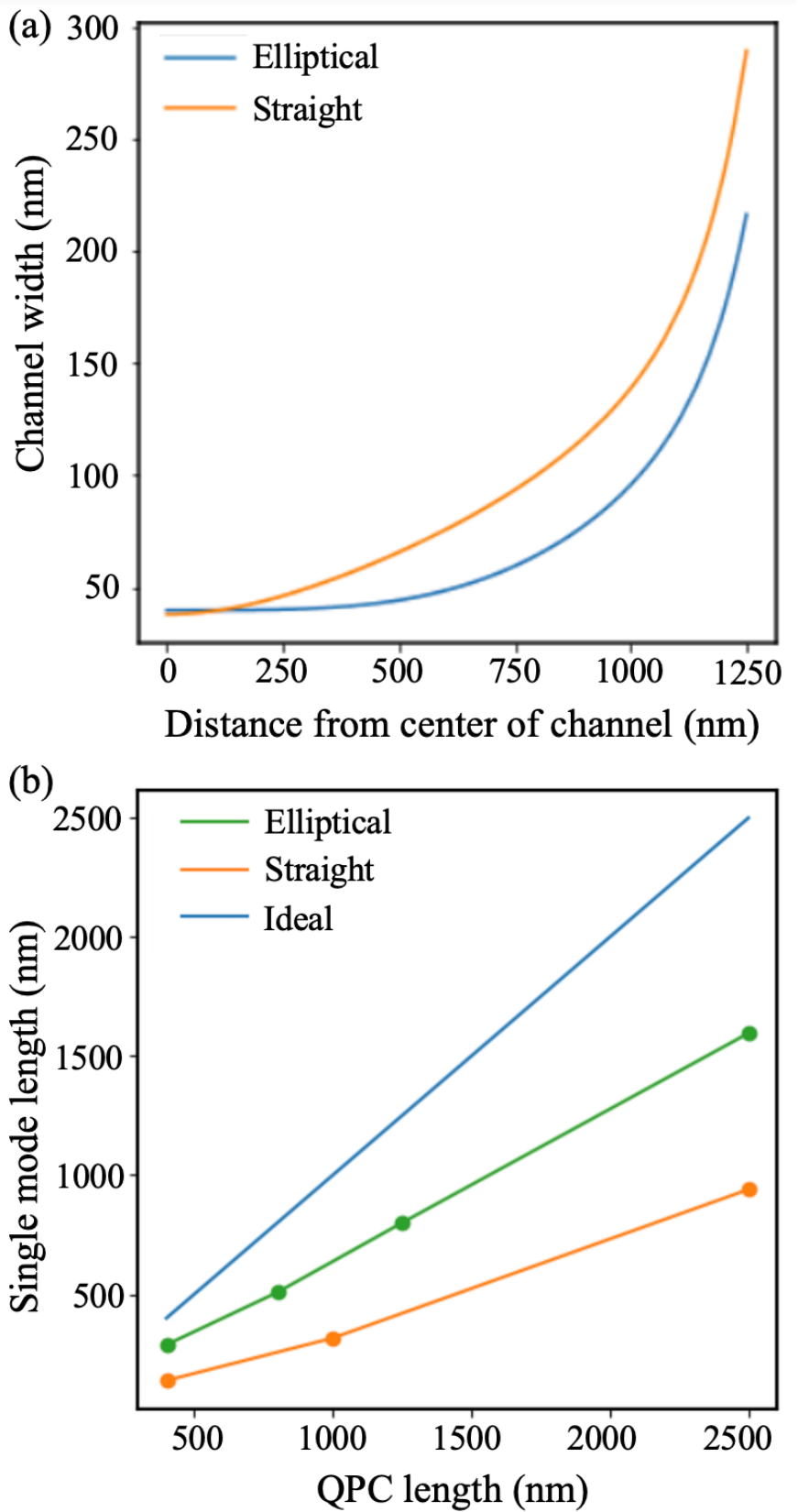}
            \caption{ 
            Characterization of the QPC optimization procedure.
            (a) Comparison of channel width variations for elliptical versus straight-gate devices of length 2500~nm. 
            The distance plotted is from the center of the channel to the lithographic edge of the device. 
            (b) Optimized single-mode lengths as a function of the lithographic gate length. Better single-mode lengths are obtained for elliptical gates, as compared to straight gates, for the same lithographic (or `ideal') gate size.
            }
            \label{fig:channel_wid_variation}
        \end{figure}

    We perform self-consistent semiclassical simulations of the QPCs using the COMSOL software package with the AC/DC Electrostatics Module \cite{COMSOL}. 
    The semiclassical model is based on a modified Thomas-Fermi approximation, as described in \cite{modifedThomasFermi}, which self-consistently determines the locally varying charge density of the 2DEG. 
    In this model, the 2DEG is assumed to form a 2D layer at the GaAs/AlGaAs interface.
    Following the approach of \cite{modifedThomasFermi}, the band offset $\Delta E_c$ is set to 0.3~eV, and we adopt an isotropic effective mass of 0.067$m_e$, as appropriate for electrons in GaAs.
    The bottom boundary of the simulation cell has the electric displacement field set to 0. We choose the Fermi level and the voltage of the global gate (which also forms the top boundary of the cell), such that the electron density in the `reservoir' portion of the device (far away from the split gates) is equal to $2\times10^{11}~\mathrm{cm}^{-2}$, while simultaneously obtaining an electrostatic potential of 0~V at the 2DEG.
    This problem is straightforwardly solved by treating the global gate and the 2DEG as the opposing plates of a parallel-plate capacitor. 

    Rectangular and elliptical split gates are also included in the COMSOL simulation geometry.
    As indicated in Fig.~\ref{fig:experimental_setup_qpc}, these gates are located directly on top of the GaAs/Al$_x$Ga$_{1-x}$As heterostructure, followed by a conformal layer of Al$_2$O$_3$, which is 15~nm thick. 
    The global gate is also patterned conformally on top of the split gates and the oxide.
    For both the rectangular and elliptical gates, we choose the voltage applied to the split gates, and the shape of the gates (in the elliptical case), such that the effective length of the 1D transport channel is maximized, following the optimization procedure described below.
    After completing these simulations, the 2D electrostatic potential at the GaAs/AlGaAs interface is output from COMSOL and input into Kwant, to perform the final conductance simulations. 

    We now describe the optimization procedure used to determine the voltages applied to the split gates.
    We note that the elliptical gate geometry has an additional optimization parameter -- the elliptical radius -- which allows the formation of longer 1D transport channels.
    Below, we also show that the channels formed under the elliptical gates are more uniform than under the rectangular gates.
    The elliptical cutouts for the gates are defined to have a major axis equal to the lithographic length of the device, while the minor axis is taken as an optimization parameter.
    After performing a given electrostatic simulation, we extract 1D linecuts from the electrostatic potential in the layer of the 2DEG, perpendicular to the transport channel (the `transverse' direction).
    Using these linecut potentials, we perform a 1D Schr\"odinger calculation to solve for the transverse eigenmodes, referred to as `subbands.'
    The Fermi level then determines which of these subbands are filled or empty.
    For each electrostatic solution, we can then define a quantity we dub the `single-mode length,' which describes the length of an unbroken channel with exactly one filled subband.
    The gate voltages, for both rectangular and elliptical gates, as well as the minor axis of the elliptical cutout, are then simultaneously optimized to achieve the longest possible single-mode length.
    The optimization procedure is repeated for each lithographic gate length.
    
    It is also interesting to define a quantity dubbed the `channel width,' which describes the classical transverse width of the channel, where the eigenenergies of the subbands cross the electrostatic confinement potential.
    In Fig.~\ref{fig:channel_wid_variation}(a), we plot this channel width as a function of the longitudinal channel position, for the optimal elliptical gate geometry, as determined above.
    Here, the solution is shown for the 2500~nm device.
    This result is also compared to the corresponding channel width for the straight-gate geometry, showing that the channel obtained for elliptical gates is significantly more uniform in width.
    In Fig.~\ref{fig:channel_wid_variation}(b), we also plot the optimal single-mode lengths for elliptical gates as compared to rectangular (`straight') gates, again obtaining significant improvements for the elliptical gates.
    Specifically, for lithographic gate lengths of 400, 800, 1250, and 2500~nm, we obtain optimized elliptical radii of 210, 85, 75, and 50~nm, respectively, and the optimized single-mode lengths are given by 288, 508, 800, and 1596~nm respectively. 
    For reference, we also plot the `ideal' mode length, corresponding to the case where the single-mode length is equal to the lithographic length of the device.

Finally, we note that the COMSOL simulations were thoroughly checked for convergence, including both the mesh density and the size of the simulation box.
The final simulations were performed with a box size of 10,000~nm in the transverse direction ($x$, perpendicular to the transport channel) and 20,000~nm in the longitudinal direction ($y$). 
    A height ($z$) of 10,000~nm was used for the thick GaAs layer. 
    To check for convergence, these three dimensions were doubled and the single-mode length was reoptimized. To determine the converged mesh size, the mesh node spacing was halved, to confirm that there was no significant change in the optimized single-mode length.

\FloatBarrier
\bibliography{references}

\end{document}